# Fully Automatic Liver Attenuation Estimation combing CNN Segmentation and Morphological Operations

Yuankai Huo [1], James G. Terry [2], Jiachen Wang [1], Sangeeta Nair [2], Thomas A. Lasko [2],
Barry I. Freedman [3], J. Jeffery Carr [2], and Bennett A. Landman [1]

[1.] Vanderbilt University, Nashville, TN, USA, 37235
[2.] Vanderbilt University Medical Center, Nashville, TN, USA, 37235
[3.] Wake Forest School of Medicine, Winston-Salem, NC, USA, 27157

## ABSTRACT

**Purpose:** Manually tracing regions of interest (ROIs) within the liver is the *de facto* standard method for measuring liver attenuation on computed tomography (CT) in diagnosing nonalcoholic fatty liver disease (NAFLD). However, manual tracing is resource intensive. To address these limitations and to expand the availability of a quantitative CT measure of hepatic steatosis, we propose the automatic liver attenuation ROI-based measurement (ALARM) method for automated liver attenuation estimation.

**Methods:** The ALARM method consists of two major stages: (1) deep convolutional neural network (DCNN)-based liver segmentation and (2) automated ROI extraction. First, liver segmentation was achieved using our previously developed SS-Net. Then, a single central ROI (center-ROI) and three circles ROI (periphery-ROI) were computed based on liver segmentation and morphological operations. The ALARM method is available as an open source Docker container (https://github.com/MASILab/ALARM).

**Results:** 246 subjects with 738 abdomen CT scans from the African American-Diabetes Heart Study (AA-DHS) were used for external validation (testing), independent from the training and validation cohort (100 clinically acquired CT abdominal scans). From the correlation analyses, the proposed ALARM method achieved Pearson correlations = 0.94 with manual estimation on liver attenuation estimations. When evaluating the ALARM method for detection of nonalcoholic fatty liver disease (NAFLD) using the traditional cut point of <40 HU, the center-ROI achieved substantial agreements (Kappa=0.79) with manual estimation, while the periphery-ROI method achieved "excellent" agreement (Kappa=0.88) with manual estimation. The automated ALARM method had reduced variability compared to manual measurements as indicated by a smaller standard deviation.

**Conclusions:** We propose a fully automated liver attenuation estimation method termed ALARM by combining DCNN and morphological operations, which achieved "excellent" agreement with manual estimation for fatty liver detection. The entire pipeline is implemented as a Docker container which enables users to achieve liver attenuation estimation in five minutes per CT exam.

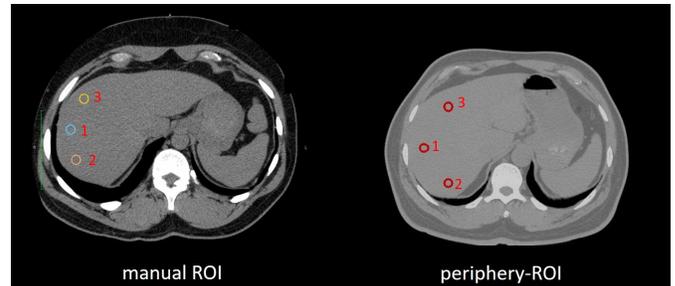

Fig. 1. The regions of interests (ROIs) of liver attenuation estimations on CT. The left panel shows an example of manual ROIs, while the right panel shows the automatic three circular ROIs (periphery -ROI) proposed in this study.

## I. INTRODUCTION

Nonalcoholic fatty liver disease (NAFLD) is the most frequent cause of liver disease worldwide and strongly associated with diabetes, metabolic syndrome, obesity liver failure, liver cancer and cardiovascular disease. (Rinella 2015, Angulo 2002) NAFLD includes a spectrum of conditions from isolated hepatic steatosis, inflammation of the liver, liver fibrosis and liver failure. NAFLD is the major cause of abnormal liver function in the United States and is expected to become the major indication for liver transplantation [1, 2]. Estimated worldwide prevalence of NAFLD ranges from 6% for diagnoses relying upon serum biomarkers of hepatic function up to 33% when diagnosed using non-invasive imaging techniques such as computed tomography (CT), ultrasound or magnetic resonance spectroscopy [3]. Moreover, patients with NAFLD are at increased risk of mortality and incidence of NAFLD is likely to rise due to its strong association with obesity and diabetes [3]. Early identification of individuals with NAFLD and nonalcoholic steatohepatitis prior to progression to fibrosis and liver failure, is considered critical to reducing the negative health consequences.

Medical imaging techniques play roles in detecting and diagnosing NAFLD in vivo. Computed tomography (CT) of the abdomen is a widely used imaging modality for a variety of conditions and quantification of hepatic steatosis has been used in both research and clinical practice. [4] Liver attenuation from CT is a non-invasive qualitative biomarker allowing measurement of liver fat content based on tissue attenuation [5, 6]. CT attenuation of normal livers typically ranges from 50 to 75 Hounsfield Units (HU) on non-contrast CT scans, while liver attenuation < 40 HU indicates moderate hepatic steatosis, with a pathologic fat content of $\geq 30\%$ [7]. CT liver attenuation is an accepted biomarker, which has been widely used for investigating hepatic steatosis (fatty liver disease) [8, 9], coronary artery risk [10, 11], diabetes [12, 13], obesity [14], and as tool for gene discovery [8].

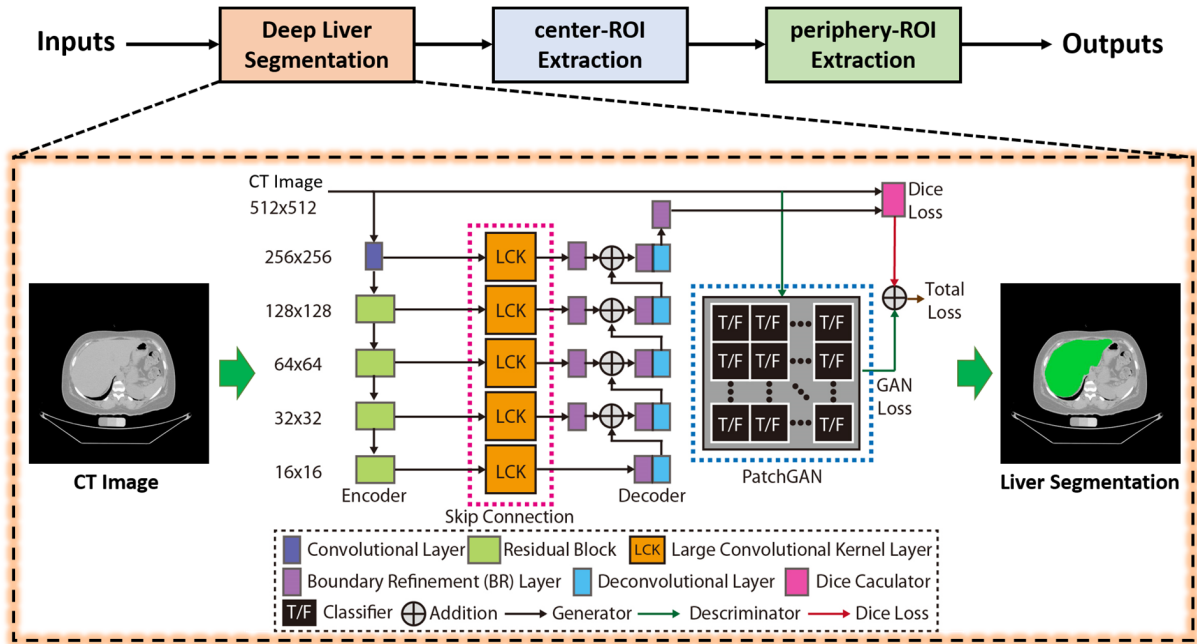

Fig. 2. The liver segmentation stage in ALARM method. Our previously developed convolutional neural network was used to segment liver. The deep network is to convert an input CT slice to a segmentation map, which consists of (1) encoder, (2) large convolutional kernel (LCK) skip connection, and (3) decoder layers. The PatchGAN is employed as the discriminator to provide additional adversarial loss for training the network with the traditional Dice loss.

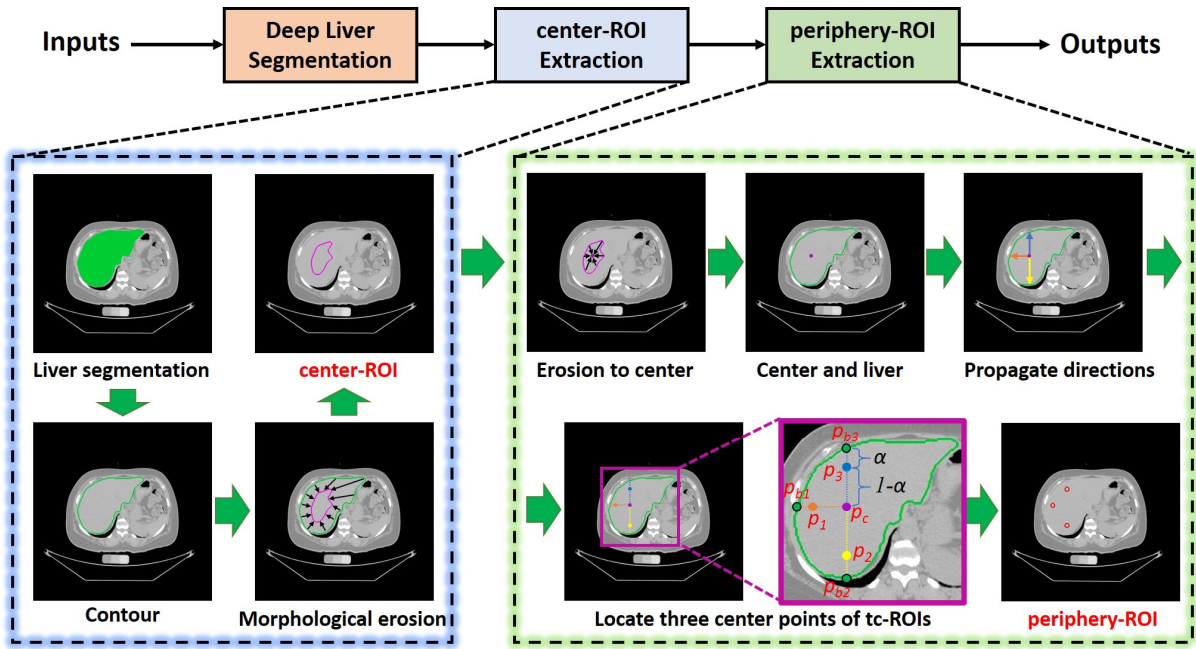

Fig. 3. The ROI extraction stage in ALARM method. Two types of automatic ROIs are obtained: (1) a single central ROI (center-ROI) and three circular ROIs (periphery-ROI).

CT based liver attenuation measurements are traditionally performed by a clinically-trained expert with many years' experience in manually tracing the liver ROIs on one or more CT slices. In the present study, the expert reader was trained by a board-certified radiologist who provides overreads of her work. This expert reader has read >1500 liver scans and has over 5 years of experience (one slice is presented Figure 1). To leverage the robustness of the ROI measurement, multiple ROIs are annotated on the CT slice, and the average HU score of each ROI from three slices is used to indicate the liver attenuation. However, manual tracing is resource and time intensive for clinical practitioners, and is not scalable for large-scale medical image data. Therefore, it is appealing to develop automatic liver attenuation estimation algorithms to alleviate the manual efforts.

Few automatic algorithms have been proposed to achieve the liver attenuation estimation on CT [15]. The straight-forward strategy is to estimate the liver attenuation from whole liver segmentation. However, (1) liver vessels and (2) non-liver tissues can be included in the whole liver segmentation, which would affect the accuracy of live attenuation estimation. Therefore, it is appealing to simulate the manual estimation protocols using fully automated image processing. In terms of manual estimation, shrinking protocol [15] and peripheral ROI protocol [7, 15, 16] are two prevalent protocols. Kullberg et al.[15] proposed an automatic liver attenuation estimation method to simulate the shrinking protocol, which shrinks the liver segmentation within the liver border to avoid non-liver tissues. However, the liver vessels are not excluded from this protocol, which would lead to higher liver attenuation values since the vessels typically have higher HU values. To further reduce the affect from vessels, the peripheral ROI protocol [7, 15, 16] has been widely used in clinical studies by manually placing several small ROIs (typically three) in the liver. Compare with single ROI, three ROIs capture the heterogeneity in the liver better, which enable more robust estimation by averaging values from different ROIs. Unfortunately, the fully automated peripheral ROI based liver attenuation estimation method has not been proposed and evaluated. Moreover, to the best of our knowledge, no previous methods have been proposed to perform peripheral ROI based liver attenuation estimation using deep learning and morphological operations.

In this manuscript, we propose the Automatic Liver Attenuation ROI-based Measurement (ALARM) pipeline, which has been developed on 100 CT scans with liver segmentation and tested on 738 additional CT scans from different populations and CT instruments used in clinical practice. The proposed pipeline simulates the ROI based manual liver attenuation labeling protocol by obtaining three liver ROIs on a non-contrast CT scan. Briefly, our previously proposed deep convolutional neural network (DCNN) based abdominal organ segmentation method [17] was employed to segment the liver. Next, the morphological operations were included to identify an ROI in the center of the liver (center-ROI) and set of three circular ROIs in the periphery of the liver (periphery-ROI) upon the liver segmentation. The entire ALARM pipeline has been implemented as a Docker container [18], which enables the users to achieve the automatic liver attenuation estimations from an abdominal CT scan by using one line of command. The ALARM pipeline can be downloaded from (https://github.com/MASILab/ALARM). To our knowledge, the proposed ALARM pipeline is the first open-source work that performs automatic ROI-based liver attenuation estimation by combining deep convolutional neural network (DCNN) and morphological operations.

II. METHODS

The end-to-end framework of the proposed ALARM pipeline is presented in Figure 2 and 3. The ALARM pipeline consists of two sections: liver segmentation and morphological operations. The liver segmentation was implemented using deep convolutional neural network described in our previous work [17].

*A. Manual Liver Attenuation Estimation Protocol*

The manual liver attenuation ROIs are traced by clinical experts, on abdominal CT scans with slice thickness of 2.5 mm and 50 field of view (FOV) ("Scan Series 1" in Figure 4). Then, a single axial view at central liver location is selected by the annotator. Next, three circles (each ~ 100 $mm^2$) are placed over the peripheral are of the right lobe of the liver (shown in Figure 1). The slice selection and circle drawing rely on the annotator's experience in clinical practice, analysts were instructed to avoid vessels, hepatic pathology and artifacts when feasible.

*B. Whole Liver Segmentation using Deep Convolutional Neural Network*

The first step in the ALARM pipeline is to obtain whole liver segmentation. The previously proposed SS-Net [17] deep convolutional neural network was employed to achieve whole liver segmentation (Figure 2). This network was originally designed for spleen MRI segmentation and was adapted to CT liver segmentation in this work. Briefly, the SS-Net is a 2D abdominal organ segmentation network consisting of both a generator and discriminator (Figure 2). The generator is to convert an input CT slice to a segmentation map, which consists of (1) encoder, (2) skip connection, and (3) decoder layers. The encoder consists of residual blocks (in green), which are employed from the canonical residual network (ResNet) [19], while the skip connection (in orange) and decoder (in purple and blue) are employed from the global convolutional network (GCN) [20]. The PatchGAN [21] is used as the discriminator to provide additional adversarial loss for training the generator. The SS-Net is trained by axial view images from 3D scans. However, the clinically acquired DICOM images are not always axial acquisition (e.g., axial, coronal, and sagittal). Therefore, to be compatible with different acquisitions, we first perform DICOM to NIFTI conversion to reconstruct the 3D volumes. Then, each axial slice in a 3D CT volume was resized to 256×256 image for both training and testing purposes. The details of preprocessing and training have previously been reported [17]. Briefly, learning rate = 0.00001, optimization method = Adam, loss function = Dice loss and Adversarial loss, and batch size = 32. After deep segmentation, the output 256×256 two-dimensional axial view segmentations were resampled to the original axial image resolution using nearest neighbor interpolation. Then, the 2D slices were stacked to 3D segmentation volumes in the original image space.

*C. Central Liver Measurement (center-ROI)*

Due to the limitation of the segmentation accuracy, the segmentation could be over-segmented or under-segmented at the boundaries of liver tissue (Figure 4b). Therefore, it would be inaccurate to get liver attenuation directly from the whole liver segmentation. To address this issue, we performed morphological erosion on liver segmentation to ensure the eroded segmentation is located within the liver. We proposed the first type of automatic liver attenuation ROI in the central region, termed center-ROI, which is an eroded segmentation based on the whole liver segmentation. The process of creating the center-ROI is shown in Figure 3, where a series of morphological erosion operations were conducted upon the whole liver segmentations. Briefly, we first resample the 3D liver binary segmentation to 1mm isotropic volume, and perform 3D erosion morphological operations until the remaining volume size <= 1000 mm$^3$. The threshold 1000 mm$^3$ is empirically chosen as an order of magnitude less than the standard liver size. The 3×3×3 kernel of erosion is a 3D "diamond" shape, where indices [1,1,0] [0,1,1] [1,0,1] [1,1,1] [1,2,1] [2,1,1] [1,1,2] are set to one while the remaining are set to zero. Then, the 3D region within the center liver ROI is defined as center-ROI, and the corresponding liver attenuation is calculated by the mean HU score within the center-ROI.

### D. Peripheral Liver ROI Measurement (periphery -ROI)

The center-ROI reflects the mean HU scores within the central liver. The central region of the liver includes the large vascular structures, specifically the portal and hepatic veins which would not be representative of hepatic fat content. Next, we perform additional operations beyond the center-ROI. The new method extracts periphery-ROI, replicating the manual process and thus avoiding the major vascular structures located in the central portion of the liver.

Since the manual ROIs are annotated on a 2D central axial slice in liver, we simulate that procedure using automated pipeline. Briefly, we continue performing morphological erosion operations on center-ROI until the volume size equals to 0. The smallest ROI from the previous non-zero volume iteration was used to locate the central point. Then, the central point $p_c = [x_c, y_c, z_c]$ is defined as the coordinate of the centroid of the small 3D volume, in which $z_c$ indicates the index of central axial slice to measure the periphery-ROI. From $p_c$, we draw three lines on posterior, lateral, and anterior direction to get the $p_{b1}, p_{b2}$, and $p_{b3}$ at the boundary of liver segmentation. $p_{b1} = [x_c, y_1, z_c], p_{b2} = [x_2, y_c, z_c]$, and $p_{b3} = [x_c, y_c, z_3]$. Next, we define the proportional coefficient $\alpha$ to obtain the three periphery-ROI center locations $p_1, p_2$, and $p_3$, whole coordinates are calculated by

$$p_1 = [x_c, y_c - (y_c - y_1)\alpha, z_c]$$
$$p_2 = [x_c - (x_c - x_1)(1 - \alpha), y_c, z_c] \quad (1)$$
$$p_3 = [x_c, y_c + (y_3 - y_c)(1 - \alpha), z_c]$$

For all the experiences in this study, the coefficients were empirically set as $\alpha = 1/3$ to (1) be tolerant to the imperfect liver segmentation, and (2) not be too close to the center point $p_c$. The ablation study of the proportional coefficient $\alpha$ is provided in Figure S2 in supplementary materials. Using $p_1, p_2$, and $p_3$ as centers, three circles of periphery-ROI are obtained, whose radius is empirically set to 7mm (area ≈150 mm$^2$). The mean HU score with each periphery-ROI is defined as liver attenuation score.

### E. ALARM Docker

The ALARM method consists of a variety of image processing algorithms (e.g., image preprocessing, deep segmentation, ROI extraction, and visualization). It would be time and resource intensive to deploy for users who lack a background in programming. Therefore, we integrated all steps in a single Docker container [18] as an "end-to-end" solution, called "ALARM Docker". The inputs of the Docker are DICOM format images from a single abdomen CT scan. The outputs of the ALARM Docker are the liver attenuation ROIs and measurements. As a result, the ALARM method can be deployed on abdominal CT scans using only one line of command. The ALARM Docker and example data from [22] is available at (https://github.com/MASILab/ALARM) with the details in the Supplementary Materials.

## III. EXPERIMENTS

### A. Data

One hundred clinically-acquired abdominal CT scans [22] with manual liver segmentation were used as training and validation data, while another independent cohort with 246 subjects from the African American-Diabetes Heart Study (AA-DHS) [23, 24] were used as external validation (testing) data. For each of the 246 AA-DHS study participants included in the external validation (testing), three abdomen series were acquired each differing only in slice thickness or FOV as shown in Figure 5. This cohort is only used in testing and independent to the training since it is excluded from training or tuning the deep learning parameters.

The 100 training and validation CT abdominal scans with liver segmentations were acquired during portal venous contrast phase with volume size $512 \times 512 \times 33$ to $512 \times 512 \times 158$. The in-plane resolution varies from $0.54 \times 0.54$ mm$^2$ to $0.98 \times 0.98$ mm$^2$, while the slice thickness ranges from 1.5 mm to 7.0 mm (details can be found in [22]). For AA-DHS scans [23, 24], each subject has three different scan series (Figure 5). In scan series 1, the field of view (FOV) is 50 mm while the slice thickness is 2.5 mm. In scan series 2, the FOV is 35 mm while the slice thickness is 2.5 mm. In scan series 3, the FOV is 35 mm while the slice thickness is 1.25 mm. For three scan series, the in-plane size is $512 \times 512$. The manual annotations were performed on scan series 1.

### B. Experimental Design

For the liver segmentation using the method in § section 2.2, 75 abdomen CT scans were randomly picked as training while the remaining 25 were used as internal validation. The 738 AA-DHS scans were not used in the training and validation; therefore, the parameters were only tuned on 100 training and validation cohort. After obtaining the whole liver segmentations, the center-ROI and periphery-ROI are computed based on the ALARM pipeline. The entire ALARM pipeline was deployed on all 738 AA-DHS scans, which took ~5 minutes for each scan using a typical workstation with Intel Xeon ES-2630 V4 2.2 GHz CPU, 32G memory, and NVIDIA Titan GPU (12 GB memory) and CUDA 8.0.

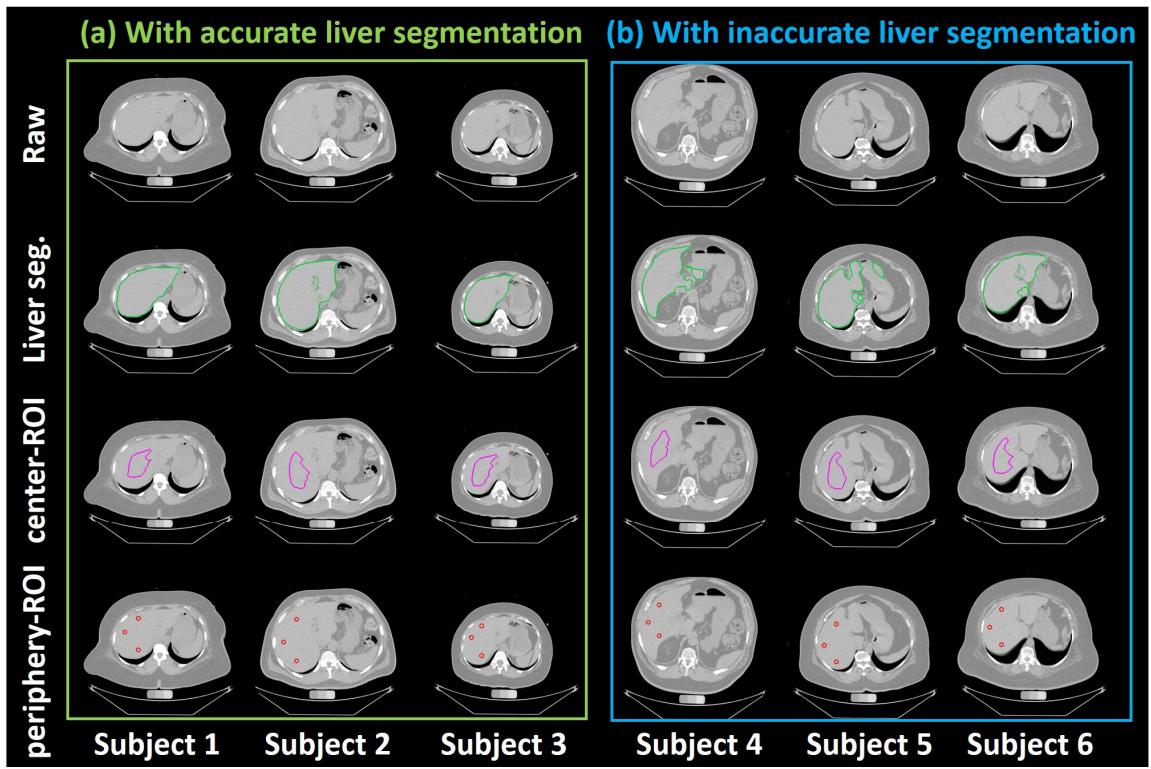

Fig. 4. Qualitative visualizations of raw abdomen CT scan, liver segmentation, center-ROI and periphery-ROI from ALARM pipeline (a) shows the results of three subjects from accurate liver segmentation, while (b) presents the results of three subjects with the inaccurate liver segmentation. The first row indicates the raw input CT scan. The second row shows the liver segmentation from the deep learning segmentation. The third row shows the central-ROI, while the fourth row shows the periphery-ROI. The ROI based liver attenuation method is able to tolerate imperfect whole liver segmentation after performing morphological operations.

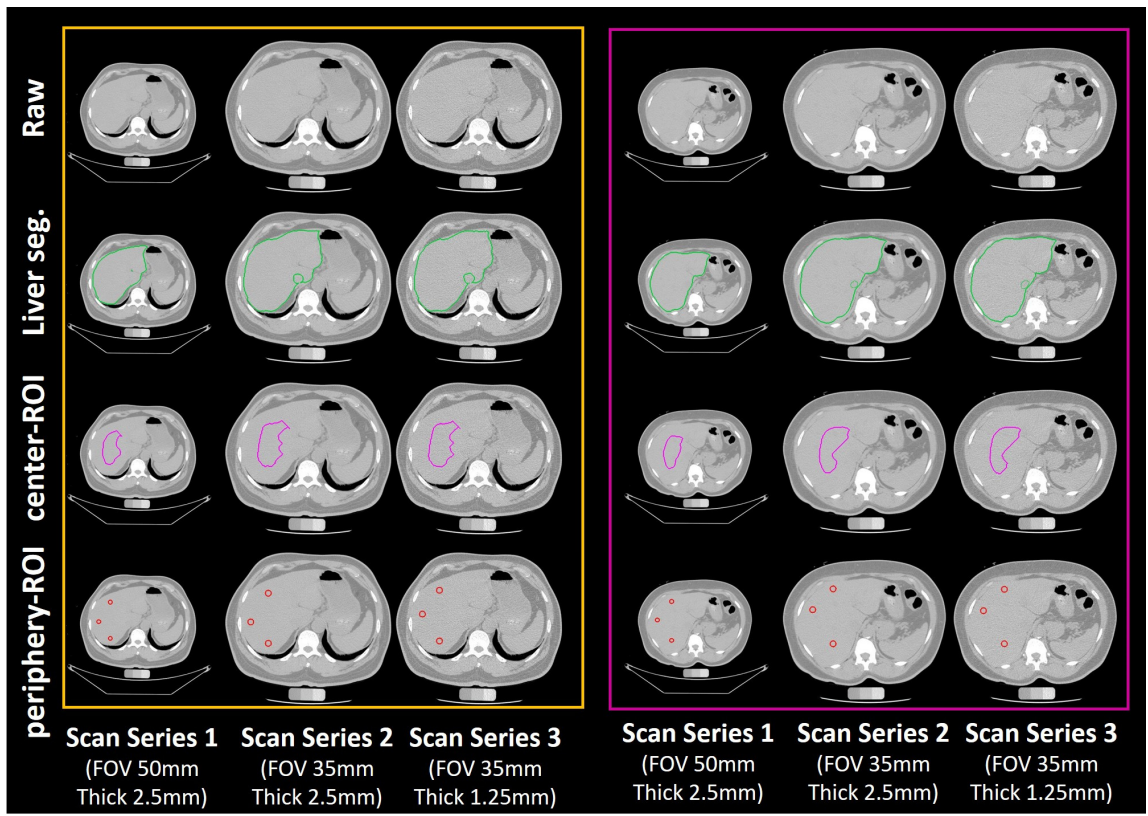

Fig. 5 Qualitative visualizations of raw abdomen CT scan, liver segmentation, center-ROI and periphery-ROI from ALARM pipeline. The left panel shows the results for one subject from three scanning protocols 1 to 3. The right panel presents the results for another subject from three scanning protocols. The first row indicates the raw input CT scan. The second row shows the liver segmentation from the deep learning segmentation. The third row shows the central-ROI, while the fourth row shows the periphery-ROI. The ROI based liver attenuation method is able to deployed on different imaging protocols.

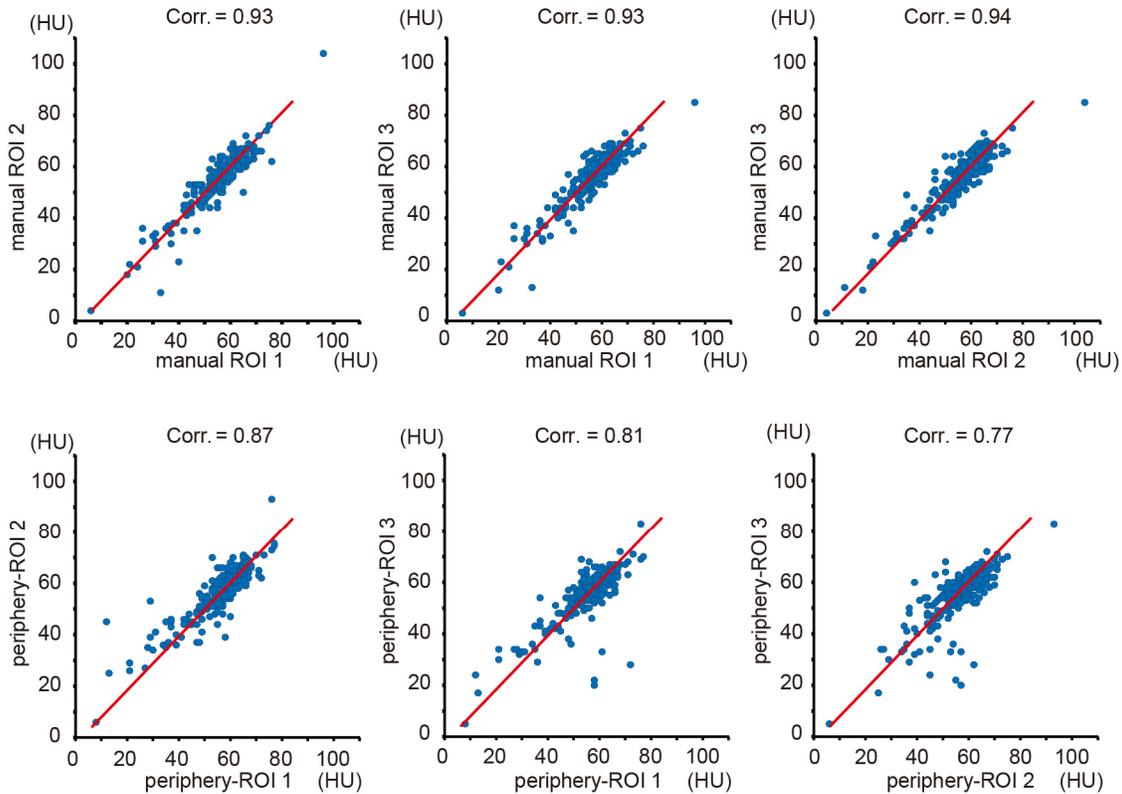

Fig. 6. The upper panel shows the scatter plots as well as the Pearson correlation between three manual ROIs, while the lower panel shows the results between three circular ROIs in periphery-POI. The blue dots indicate the Hounsfield Unit (HU), and the red lines are the linear regression results.

## IV. STATISTICAL ANALYSES

Summary measures including means, standard deviations (SD), medians, ranges, and other measures of distribution were used to assess performance of center-ROI and periphery-ROI and manual methods. Also presented are scatterplots and Pearson correlation coefficients that show relationships within methods and between methods across the range of liver values encountered. We also present Bland-Altman plot assessing the limits of agreement for our final model comparing the mean of three ROIs measured by the periphery-ROI and manual methods [25]. As liver attenuation values <40 HU are considered indicative of moderate to severe steatosis, we also evaluated the methods on the basis of number of mean attenuations falling below or at or above the 40 HU threshold [6, 26, 27]. Predictive agreement for identifying <40 HU was tested using kappa (K) and 95% confidence intervals (95% CI) [28, 29] and suggested scales for assessing agreement are presented [22]. All analyses were performed using STATA 15.1 (StataCorp LLC, College Station, TX, USA) and Bland-Altman and scatterplots were craested using GraphPad Prism 6.05 (GraphPad Software, Inc., LaJolla, CA. USA).

## V. RESULTS

Figures 4 and 5 present the qualitative results from the ALARM pipeline. Figure 4 shows the segmentation and ROI results of six subjects. The center-ROI and periphery-ROI results for the three subjects in the left panel are generated when the liver segmentations are accurate. The center-ROI and periphery-ROI results for the three subjects in the right panel are achieved when the liver segmentations are inaccurate. Results demonstrate that the proposed method is able to identify well placed periphery-ROI even when the liver segmentation is not accurate. Figure 5 shows the segmentation and ROI results of two subjects. For each subject, the liver segmentation, center-ROI, and periphery-ROI results are presented for the three different scan series. Note that the manual liver segmentation results are not shown in Figure 4 and 5 since those patients are from AA-DHS testing cohort, in which we do not have manual liver segmentation results.

### A. Segmentation Performance

The liver segmentation performance of the SS-Net is evaluated on 25 internal validation abdominal CT scans with manual liver annotation. The segmentation performance with median DSC values of SS-Net on the 100 epochs is provided in Figure S1 in the Supplementary Materials. From the validation results, the segmentation model with epoch=85 is chosen for the remaining morphological operations, which achieves median DSC = 0.942, mean DSC = 0.934, standard deviation = 0.021 on 25 internal validation cohorts.

### B. Correlations of Manual ROI and Periphery-ROI

Three ROIs are identified for both manual ROI delineation and periphery-ROI extractions. Figure 6 shows the Pearson correlation of liver attenuation estimation across three ROIs. The upper panel shows the cross ROI correlations for manual delineation, while the lower panel shows the cross ROI correlations for periphery-ROI estimations. From the results, the correlations between manual ROIs are higher than periphery-ROIs.

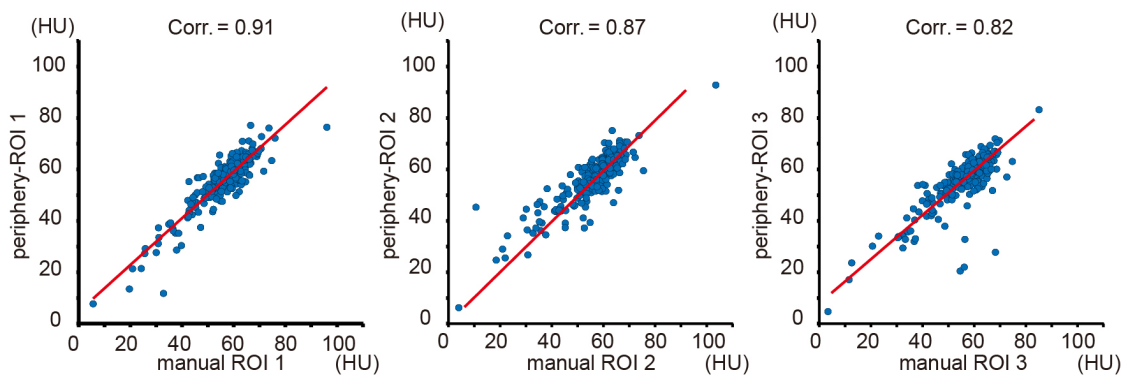

Fig. 7. The correlation maps between each manual ROI and each periphery-ROI. The blue dots indicate the Hounsfield Unit (HU), and the red lines are the linear regression results.

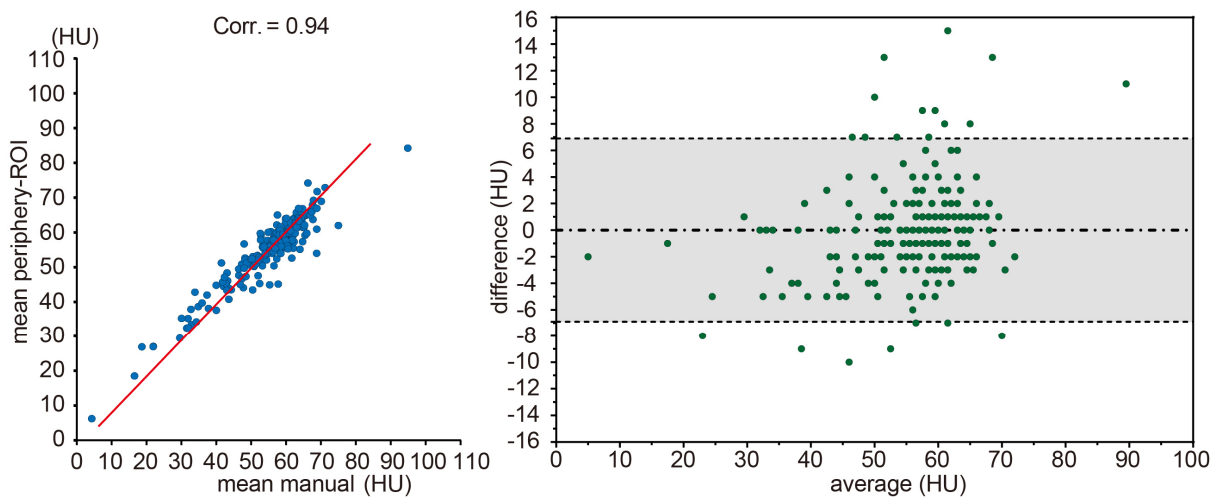

Fig. 8. The left panel shows the correlations between mean periphery-ROI and mean manual ROI. The blue dots indicate the Hounsfield Unit (HU), and the red lines are the linear regression results. The right panel shows the Bland-Altman plot between mean periphery-ROI vs. mean manual ROI. The gray area indicates the 95% confidence interval.

Then, one-to-one ROI wise correlations between each manual ROI and periphery-ROI is presented in Figure 7. From Figure 6 and 7, The periphery-ROI has larger inter-ROI variations. Therefore, to reduce the variations for automatic liver attenuation estimation, the mean HU scores across three ROIs for both manual and automatic methods is used in this study, whose Bland-Altman plot [25] are shown in Figure 8. The left panel in Figure 8 shows that we achieve Pearson correlation = 0.94 between mean periphery-ROI and mean manual ROI for liver attenuation estimation. The right panel in Figure 8 presents the Bland-Altman plot with 95% confidence interval. The x-axis indicates the average of mean periphery-ROI and mean manual ROI, while the y-axis indicates the differences by subtracting mean periphery-ROI from mean manual ROI. The upper and lower dash lines indicate the 95% confidence interval (6.916 and -6.938), while the middle dash line shows the bias (-0.011).

### C. Quantitative Results on Fatty Liver Detection

The quantitative results have been presented in Table 1 and 2. Table 1 shows the HU scores along with mean, SD and range for center-ROI and periphery-ROI across three scan series (1-3) measured by ALARM method as well as HU scores for series 1 by manual method are provided in Table 1. The row "HU<40" in Table 1 indicates the estimation number of subjects with fatty liver based on Hounsfield units (HU<40). On the same scan series 1, 16 subjects were detected with fatty liver using center-ROI, while 17 subjects were detected with fatty liver using periphery-ROI. Using manual measurement, 19 subjects were identified with fatty liver. Agreement between automatic and manual measurements are evaluated using the K statistic [28]. The K statistic on inter-rater agreement between manual and the proposed ALARM methods on fatty liver detection (HU<40) are presented in Table 2. The K values are all > 0.7, which indicate substantial agreement between manual and the ALARM methods. The fatty liver detection based on the periphery-ROI ALARM method agreement with manual estimation (K = 0.88, p<0.0001) is considered to be "almost perfect" [30]. When applying ALARM method on different series (Series 2 and 3), the center-ROI achieved K = 0.82 and 0.84 with manual detection for Series 2 and 3. The periphery-ROI achieved K = 0.91 and 0.72 with manual detection for Series 2 and 3. The sensitivity and specificity results of the detection performance are provided in Table 2 as well.

TABLE I
THE LIVER ATTENUATION MEASUREMENTS OF CENTER-ROI AND MEAN PERIPHERY-ROI FOR DIFFERENT SEQUENCE PROTOCOL

|  | center-ROI | | | periphery-ROI | | | Manual |
|---|---|---|---|---|---|---|---|
|  | **Series 1** | Series 2 | Series 3 | **Series 1** | Series 2 | Series 3 | **Series 1** |
| Mean, HU | **56.6** | 56.7 | 56.9 | **55.9** | 56.7 | 56.9 | **55.9** |
| SD, HU | **9.3** | 9.3 | 9.2 | **9.4** | 9.3 | 9.2 | **10.4** |
| Median, HU | **58.5** | 58.6 | 58.6 | **57.8** | 58.6 | 58.6 | **58.0** |
| Minimum, HU | **7.2** | 8.7 | 8.5 | **6.2** | 8.7 | 8.5 | **4.3** |
| Maximum, HU | **88.9** | 89.9 | 97.7 | **84.3** | 89.9 | 97.7 | **94.9** |
| 25 percent, HU | **53.9** | 54.1 | 54.2 | **52.4** | 54.1 | 54.2 | **52.3** |
| 75 percent HU | **62.5** | 62.3 | 62.5 | **61.6** | 62.3 | 62.5 | **62.5** |
| HU<40, n(%) | **16 (6.4%)** | 17 (6.8%) | 14 (5.6%) | **17 (6.8%)** | 16 (6.4%) | 19 (7.5%) | **19 (7.5%)** |

\* Scan Series 1: 2.5 mm, 50 FOV; Scan Series 2: 2.5 mm, 35 FOV; Scan Series 3: 1.25 mm, 35 FOV

TABLE II
AGREEMENT OF CENTER-ROI AND PERIPHERY-ROI WITH MANUAL METHOD FOR DETECTION OF LIVER ATTENUATION <40 HU

|  | ALARM center-ROI vs. Manual | | | ALARM periphery-ROI vs. Manual | | |
|---|---|---|---|---|---|---|
|  | **Series 1** | Series 2 | Series 3 | **Series 1** | Series 2 | Series 3 |
| Kappa | **0.79** | 0.82 | 0.84 | **0.88** | 0.91 | 0.72 |
| 95% CI | **0.63 – 0.94** | 0.68 – 0.96 | 0.70 – 0.98 | **0.77 – 0.99** | 0.81 – 0.99 | 0.55 – 0.88 |
| p | **< 0.0001** | < 0.0001 | < 0.0001 | **< 0.0001** | < 0.0001 | < 0.0001 |
| Agreement | **97.2%** | 97.6% | 98.0% | **98.4%** | 98.8% | 96.0% |
| Sensitivity | **73.7%** | 79.0% | 73.7% | **84.2%** | 84.2% | 73.7% |
| Specificity | **99.1%** | 99.2% | 100.0% | **99.6%** | 100.0% | 97.9% |

\* Kappa agreement ranges: 0.21-0.40, fair; 0.41-60, moderate; 0.61-0.80, substantial; >081, almost perfect based on Landis and Koch (1977).

## VI. DISCUSSION

In the present study, we evaluated the use of the ALARM method for automated measurement of liver attenuation and prevalence of NAFLD, an important marker of future cardiovascular risk. To the best of our knowledge, the proposed ALARM pipeline is the first open-source pipeline that performs automatic ROI-based liver attenuation estimation by combining DCNN and morphological operations. ALARM method determination of liver attenuation and NAFLD detection was highly correlated with gold-standard manual tracings performed by a trained analyst. From Figure 5 and Table 1, the proposed ALARM method is able to achieve good automatic liver attenuation estimation across different scan series. The manual protocols of liver attenuation estimation can be done efficiently in clinical practice. However, it might not be a scalable method and typically not desired in large-scale imaging analysis studies, especially for the large-scale retrospective studies. Using the proposed ALARM method, we are able to achieve the liver attenuation measurements from the large-scale cohorts automatically.

The prevalence of NAFLD, which may already affect up to 33% worldwide, may be expected to only increase in the coming years secondary to the obesity epidemic. NAFLD is associated with higher risk for cardiovascular disease risk and progression of NAFLD to more serious liver disease increases risk of hepatic cancer, liver failure and the need for liver transplantation. It is thus imperative that large scale means for identifying those at risk for NAFLD be developed. CT is often used for population studies of liver attenuation and NAFLD, but these studies are limited by the resource-intensive need to train analysts to manually place liver ROIs consistently. Nevertheless, liver attenuation has been measured thousands of study participants in Framingham [31], Jackson Heart Study [32], CARDIA [33] and MESA [34] and pooled to provide evidence of the genetic components of NAFLD [8]. Thus, we developed and tested the ALARM method for liver attenuation determination potentially opening the door for measurement of fatty liver in tens of thousands of study participants or patients.

Using the proposed method, the resource consuming manual annotation is eliminated from the pipeline. Moreover, the ALARM method typically takes five minutes to perform liver attenuation estimation from a abdomen CT scan. Therefore, we are able to deploy the ALARM method on the larger scale clinical cohorts without manual efforts to understand the relationship between liver attenuation and diseases. From the results, the proposed method achieved "excellent" agreement with manual measurement on the fatty liver detection.

The present study has some limitations. One limitation of the proposed ALARM method is that the whole pipeline has not been implemented in a single "end-to-end" deep convolutional neural network. The deep convolutional neural network was only employed for liver segmentation, but not for the ROI extraction. In the future, the improvement could be to implement a single end-to-end deep network to perform multi-task learning on liver segmentation and ROI detection simultaneously [35]. Although performance was excellent, it could even be enhanced by training a landmark detection DCNN directly using labeled ROIs on >1000 subjects.

Another limitation of the proposed ALARM method is that the liver vessels were not explicitly considered when extracting the ROIs in automatic methods. Although the periphery-ROI method has been designed to mimic the ROI placement used in most manual methods and thus reduces the effects of liver vessels, the ROIs could still contain vessels. As a result, liver attenuation estimation could be affected since the HU scores are different between vessels and liver tissues and the periphery-ROI does not guarantee the elimination of the non-liver tissues (i.e. veins and arteries). In the future, the automated liver attenuation estimation methods would achieve comparable or even better performance compared with manual protocols if the non-liver tissues segmentation is introduced into the pipeline to avoid the non-liver tissue in ROI extraction.

Moreover, the present test of the ALARM method was performed in a community-based study using careful research protocols for CT acquisition and, as such, we cannot directly extrapolate its use to clinically-obtained scans and other populations including those with clinically-significant liver disease. The peripheral ROI is designed as 2D measurements to simulate the manual ROI protocol. In clinical scenarios, a single overlay image with automated 2D ROIs can be visualized for efficient quality assurance (QA), which would require fewer human efforts for completed QA the results of 3D ROIs slice-by-slice.

The 246 AA-DHS testing scans are independent to the 100 clinical acquired training and validation scans to ensure the fair external validation. Therefore, the manually traced circles and slice numbers are not included in developing the method. As a result, the slice location of the peripheral ROIs can be different from the slice location of the manual ROIs, and we do not force such consistency. The ALARM method is proposed to simulate both shrinking protocol (center-ROI) and peripheral ROI protocol (periphery-ROI) in a single fully automated pipeline. Since the periphery-ROI is achieved based on the center-ROI, the failure of center-ROI might lead to the failure of periphery-ROI (although this did not happen in the present study). ALARM method Moderate inaccuracy in whole liver segmentation (Figure 4) is tolerable and will still permit accurate center-ROI and periphery-ROI placements. However, the global failure of liver segmentation (not encountered in this study) might lead to the global failures of center-ROI and periphery-ROI. Using different segmentation methods (SS-Net and U-Net), the proposed ALARM framework achieves consistent and comparable liver attenuation performance (Figure S3, Table S1 and S2). The results demonstrate that the strategies for center-ROI and periphery-ROI are tolerant to moderately inaccurate whole liver segmentation in different scenarios (SS-Net and U-Net). Details are provided in the Supplementary Materials.

The aim of this study is to develop an automatic liver attenuation estimation method simulating human estimation protocols, by combining deep learning segmentation and morphological operations. In this study, SS-Net is employed as the deep learning segmentation method in the ALARM method, without claiming the SS-Net is an optimal solution. We have shown that the ALARM method is an open framework that allows users to adapt different segmentation approaches (i.e., the U-Net [36] in Supplementary Materials). Details of the quantitative performance comparing ALARM-U (U-Net) with ALARM (SS-Net) are provided in Supplementary Materials. Other deep learning methods might have resulted in more favorable results but are outside the scope of this work.

## VII. CONCLUSION

In this paper, a fully automated liver attenuation measurement method is proposed. This method computes liver attenuation in five minutes by incorporating deep learning and morphological operations. Liver attenuation measured using the ALARM method was highly correlated with attenuation measured manually by a clinically trained, highly experienced analyst. Moreover, when compared to the gold-standard manual method for detection of NAFLD based on the attenuation cut point of 40 HU, the proposed ALARM method produced "excellent" agreement (K = 0.88) with manual measurement. The present study suggests the ALARM method may be used to reliably measure liver attenuation and assess NAFLD prevalence in large epidemiologic studies and scan repositories potentially including tens of thousands of participants.

## VIII. ACKNOWLEDGEMENTS

This research was supported by NSF CAREER 1452485, NIH grants 5R21 EY024036, R01 EB017230 (Landman), R01 NS095291 (Dawant), R01 AR048797 (Carr) and R01 DK071891 (Freedman). This research was conducted with the support from Intramural Research Program, National Institute on Aging, NIH. This work was also supported by the National Institutes of Health in part by the National Institute of Biomedical Imaging and Bioengineering training grant T32-EB021937. This study was in part using the resources of the Advanced Computing Center for Research and Education (ACCRE) at Vanderbilt University, Nashville, TN. This project was supported in part by ViSE/VICTR VR3029 and the National Center for Research Resources, Grant UL1 RR024975-01, and is now at the National Center for Advancing Translational Sciences, Grant 2 UL1 TR000445-06. We appreciate the NIH S10 Shared Instrumentation Grant 1S10OD020154-01 (Smith), Vanderbilt IDEAS grant (Holly-Bockelmann, Walker, Meliler, Palmeri, Weller), and ACCRE's Big Data TIPs grant from Vanderbilt University. We gratefully acknowledge the support of NVIDIA Corporation with the donation of the Titan X Pascal GPU used for this research. The imaging dataset(s) used for the analysis described were obtained from ImageVU, a research resource supported by the VICTR CTSA award (ULTR000445 from NCATS/NIH), Vanderbilt University Medical Center institutional funding and Patient-Centered Outcomes Research Institute (PCORI; contract CDRN-1306-04869). The authors have no conflicts to disclose.

SUPPLEMENTARY MATERIALS

*A. Alarm Docker*

Docker (https://www.docker.com) is a containerized technique that provides an open-source and lightweight solution to deploy developed algorithms on any operating system (OS). Compared with virtual machines, Docker containers use smaller and neater capsule upon OS rather than employing hypervisors, which emulates the virtual hardware. As the GPU based computing was used in the ALARM method, the GPU extension version of Docker, NVIDIA-Docker (https://github.com/NVIDIA/nvidia-docker), was used as the containerized implementation.

The Docker was established on Ubuntu 16.04 with CUDA 8.0, MATLAB 2016a, Python 2.7, and PyTorch 0.2. The Docker has been made publicly available in our Dockerhub (https://hub.docker.com/u/masidocker), which can be deployed to local computer by calling the following command (provided with Ubuntu bash as example):

sudo docker pull masidocker/spiders:liver_attenuation_v1_0_1

Once the docker has been imported to the local computer, users are able to obtain the final output files by running a single command:

sudo nvidia-docker run -it --rm -v {input path}:/INPUTS/ -v {output path}:/OUTPUTS masidocker/spiders:liver_attenuation_v1_0_1 /extra/run_deep_wholebody.sh

*B. Ablation Study*

The quantitative performance of training (75 CT scans) and validation (25 CT scans) using the SS-Net is provided in the Figure S1. In ALARM method, the default value of the proportional coefficient $\alpha$ is 1/3. The ablation test of $\alpha = \left[\frac{1}{6}, \frac{1}{3}, \frac{1}{2}, \frac{2}{3}, \frac{5}{6}\right]$ is provided in Figure S2.

*C. Quantitative Results*

The quantitative results of replacing SS-Net with U-Net in ALARM (called "ALARM-U") are provided in the supplementary materials. The training data and training hyper-parameters to train SS-Net are kept same for training U-Net, except the batch size = 16 for U-Net due to the memory limitation. The mean HU scores across three peripheral ROIs for both manual and automatic ALARM-U methods are used in this study and the e scatterplot and Bland-Altman plot are shown in Figure S3. The left panel in Figure S3 shows the high correlation (r=0.95) between mean periphery-ROI and mean manual ROI for liver attenuation estimation. The right panel presents the Bland-Altman plot with 95% confidence interval. The x-axis indicates the average of mean periphery-ROI and mean manual ROI, while the y-axis indicates the differences by subtracting mean periphery-ROI from mean manual ROI. The detailed quantitative results of ALARM-U have been presented in Tables S1 and S2.

*D. Compare ALARM and ALARM-U*

Figure S3 shows the correlations between mean periphery-ROI and mean manual ROI for the ALARM-U method.

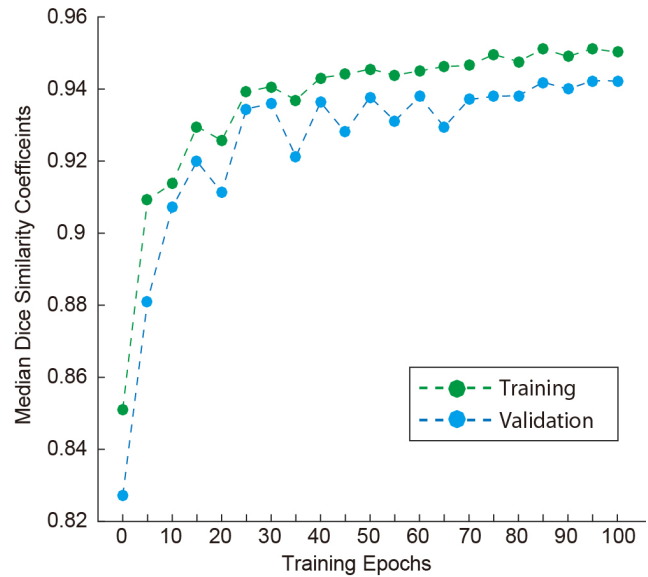

Fig. S1. Liver segmentation performance with median Dice similarity coefficient (DSC) of SS-Net across 100 training epochs is provided in the blue curve. The corresponding performance for training is provided in the red curve

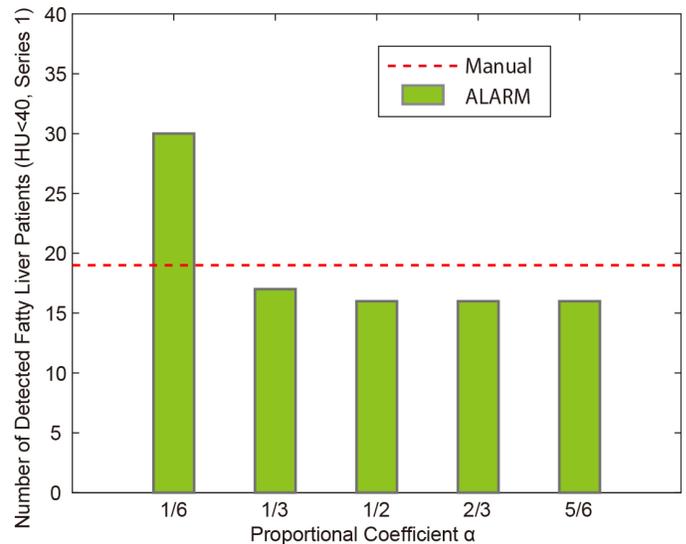

Fig. S2. The ablation study of the proportional coefficient α in fatty liver detection using series 1 scans. The green bars indicate the number of detected fatty liver cases among 246 patients in AADHS study. The red dash line shows 19 cases are detected from manual protocol. From the results, the default value α=1/3 has the closest estimation. The α=1/6 (closer to boarder) detects more false positive cases since the HU scores are lower when the ROIs cover the areas outside the liver boarder.

Compared with the ALARM in Figure 9, the ALARM-U achieves slightly higher correlation. However, the ALARM-U yields larger difference (y-axial) range in the Bland-Altman plot compared with ALARM method. Table S1 and S2 for ALARM-U present the same comparisons shown for ALARM in Tables 1 and 2. The ALARM method achieves better Kappa scores in terms of periphery-ROI on series 1. However, generally comparable liver attenuation estimation performances are achieved by employing heterogeneous deep learning segmentation methods in the ALARM pipeline.

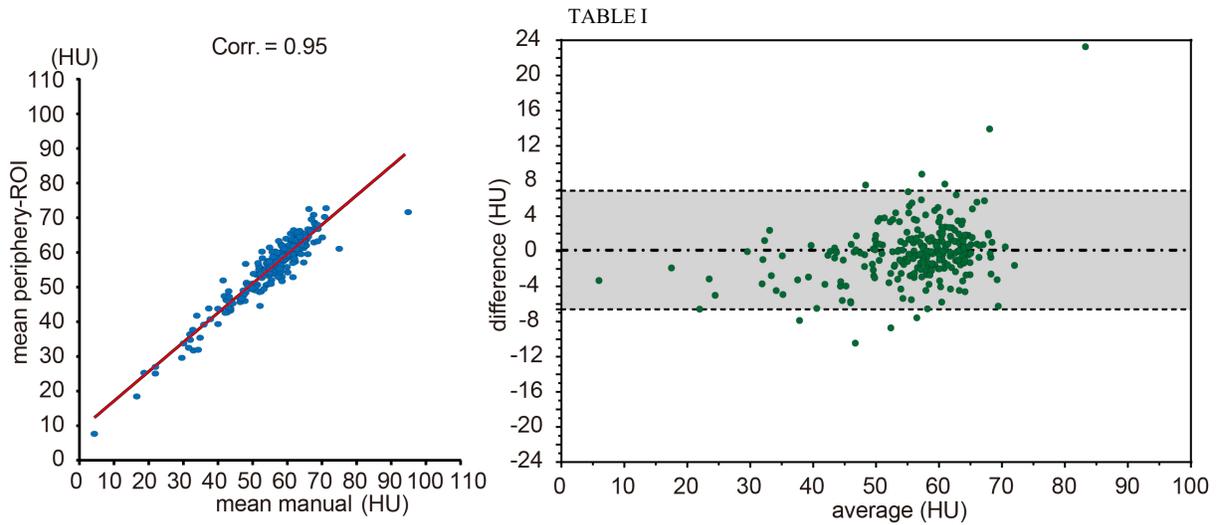

Fig. S3. The left panel shows the correlations between mean periphery-ROI and mean manual ROI. The blue dots indicate the Hounsfield Unit (HU), and the red lines are the linear regression results. The right panel shows the Bland-Altman plot between mean periphery-ROI vs. mean manual ROI. The gray area indicates the 95% confidence interval.

TABLE SI
THE LIVER ATTENUATION MEASUREMENTS OF CENTER-ROI AND MEAN PERIPHERY-ROI FOR DIFFERENT SEQUENCE PROTOCOL

|  | ALARM-U center-ROI | | | ALARM-U periphery-ROI | | | Manual |
| --- | --- | --- | --- | --- | --- | --- | --- |
|  | **Series 1** | Series 2 | Series 3 | **Series 1** | Series 2 | Series 3 | **Series 1** |
| Mean, HU | **57.2** | 57.2 | 57.2 | **56.0** | 55.9 | 56.1 | **55.9** |
| SD, HU | **9.2** | 9.6 | 9.1 | **9.4** | 9.6 | 9.5 | **10.4** |
| Median, HU | **58.9** | 58.8 | 59.0 | **58.3** | 57.9 | 58.1 | **58.0** |
| Minimum, HU | **8.6** | 10.1 | 9.2 | **7.6** | 4.2 | 8.4 | **4.3** |
| Maximum, HU | **89.7** | 118.4 | 92.0 | **72.8** | 106.9 | 79.7 | **94.9** |
| 25 percent, HU | **54.2** | 54.3 | 54.3 | **52.7** | 52.7 | 52.6 | **52.3** |
| 75 percent HU | **63.0** | 62.8 | 63.0 | **62.1** | 61.9 | 62.7 | **62.5** |
| HU<40, n(%) | **16 (6.3%)** | 15 (5.9%) | 15 (5.9%) | **17 (6.6%)** | 18 (7.1%) | 15 (5.9%) | **19 (7.5%)** |

\* Scan Series 1: 2.5 mm, 50 FOV; Scan Series 2: 2.5 mm, 35 FOV; Scan Series 3: 1.25 mm, 35 FOV

TABLE SII
AGREEMENT OF CENTER-ROI AND PERIPHERY-ROI WITH MANUAL METHOD FOR DETECTION OF LIVER ATTENUATION <40 HU

|  | ALARM-U center-ROI vs. Manual | | | ALARM-U periphery-ROI vs. Manual | | |
| --- | --- | --- | --- | --- | --- | --- |
|  | **Series 1** | Series 2 | Series 3 | **Series 1** | Series 2 | Series 3 |
| Kappa | **0.81** | 0.77 | 0.84 | **0.85** | 0.88 | 0.75 |
| 95% CI | **0.66 – 0.96** | 0.61 – 0.94 | 0.70 – 0.98 | **0.72 – 0.98** | 0.77 – 0.99 | 0.58 – 0.92 |
| p | **< 0.0001** | < 0.0001 | < 0.0001 | **< 0.0001** | < 0.0001 | < 0.0001 |
| Agreement | **97.7%** | 97.2% | 98.0% | **98.0%** | 98.4% | 96.9% |
| Sensitivity | **73.7%** | 68.4% | 73.7% | **79.0%** | 84.2% | 68.4% |
| Specificity | **99.6%** | 99.6% | 100.0% | **99.6%** | 99.6% | 99.2% |

\* Kappa agreement ranges: 0.21-0.40, fair; 0.41-60, moderate; 0.61-0.80, substantial; >081, almost perfect based on Landis and Koch (1977).